\documentclass[preprint]{revtex4}

\usepackage{epsfig}

\def\bea{\begin{eqnarray}}
\def\eea{\end{eqnarray}}
\def\bec{\begin{center}}
\def\ec{\end{center}}

\def\ler{\lesssim}
\def\gtr{\gtrsim}
\def\beq{\begin{equation}}
\def\eeq{\end{equation}}

\def\ler{\lesssim}
\def\gtr{\gtrsim}

\def\p{\partial}
\def\f{\frac}

\def\f#1#2{\frac{#1}{#2}}
\def\p{\partial}

\def\l{\left}
\def\r{\right}

\begin{document}
\draft
\tighten
\preprint{KAIST-TH 03/14, TU-706}
\title{\large \bf Electroweak Symmetry Breaking
in Supersymmetric Gauge-Higgs Unification Models}
\author{ 
Kiwoon Choi,$^a$\footnote{kchoi@hep.kaist.ac.kr} 
Naoyuki Haba,$^b$\footnote{haba@ias.tokushima-u.ac.jp}
Kwang-Sik Jeong,$^a$\footnote{ksjeong@hep.kaist.ac.kr}\\
Ken-ichi Okumura,$^a$\footnote{okumura@hep.kaist.ac.kr}
Yasuhiro Shimizu,$^c$\footnote{shimizu@tuhep.phys.tohoku.ac.jp} 
Masahiro Yamaguchi\phantom{,}$^c$\footnote{yama@tuhep.phys.tohoku.ac.jp}
}
\address{
$^a$Department of Physics, Korea Advanced Institute of Science
and Technology, Daejeon 305-701, Korea\\
$^b$Institute of Theoretical Physics, University of Tokushima, Tokushima
770-8502, Japan\\
$^c$Department of Physics, Tohoku University,
 Sendai 980-8578, Japan}
\date{\today}
%\maketitle
%%%%%%%%%%%%%%%%%%%%%%%%%%%%%%%%%%%%%%%%%%%%%%%%%%%%%%
\begin{abstract}
%%%%%%%%%%%%%%%%%%%%%%%%%%%%%%%%%%%%%%%%%%%%%%%%%%%%%%
We examine the Higgs mass parameters
and electroweak symmetry breaking
in supersymmetric orbifold field theories
in which the 4-dimensional Higgs fields originate from
higher-dimensional gauge supermultiplets.
It is noted that such gauge-Higgs unification leads to
a specific boundary condition on the Higgs mass parameters
at the compactification scale, which is independent
of the details of supersymmetry breaking mechanism.
With this boundary condition, phenomenologically viable
parameter space of the model is severely  constrained
by the condition of electroweak symmetry breaking
for supersymmetry breaking scenarios which can be realized naturally
in orbifold field theories.
For instance, if it is assumed that the 4-dimensional effective theory
is the minimal supersymmetric standard model with
supersymmetry breaking parameters induced mainly by the
Scherk-Schwarz mechanism,
a correct electroweak symmetry breaking can not be achieved
for reasonable range of parameters of the model,  even when
one includes additional contributions to the Higgs mass parameters from
the auxiliary component of 4-dimensional conformal compensator.
However if there exists a supersymmetry breaking mediated by
brane superfield, sizable portion of the parameter space can give
a correct electroweak symmetry breaking.

%This implies that there should exist
%another source of supersymmetry breaking, e.g. the auxiliary component
%of a brane superfield.

%%%%%%%%%%%%%%%%%%%%%%%%%%%%%%%%%%%%%%%%%%%%%%%%%%%%%%%
\end{abstract}
%%%%%%%%%%%%%%%%%%%%%%%%%%%%%%%%%%%%%%%%%%%%%%%%%%%%%%%
\pacs{}
\maketitle

\section{introduction}

It has been noticed that theories with extra dimension
can provide an elegant mechanism to generate various hierarchical
structures in 4-dimensional (4D) physics, e.g.
the scale hierarchy $M_W/M_{Pl}\approx 10^{-16}$ \cite{add}
or the doublet-triplet splitting in grand unified
theories \cite{TD-splitting-etc}.
Another interesting possibility with
extra dimension is that 4D Higgs fields originate from
higher-dimensional gauge fields, unifying Higgs fields
with gauge fields \cite{GH-old-1,GH-1}.
However, constructing
a realistic model of gauge-Higgs unification
is non-trivial because of the difficulties to obtain
the Higgs quartic coupling and a realistic form of Yukawa couplings
\cite{GH-now-1,GH-now-2}.
Recently it has been pointed out that the idea of
gauge-Higgs unification can be implemented successfully
within the framework of supersymmetric orbifold
field theories \cite{BN,GH-now-3,GH-now-4} in which the Higgs quartic
%couplings are given by the usual $D$-term potential
couplings can be given by the usual $D$-term potential
of $SU(2)\times U(1)$, and also a realistic form of
Yukawa couplings can be obtained through
the quasi-localization of bulk fermions and the mixings
with brane fermions.

In this paper, we examine the Higgs mass parameters
and the resulting electroweak symmetry breaking
in supersymmetric gauge-Higgs unification models.
To be specific, we will focus on
5D models, however some of our results are valid in more general cases.
One model-independent prediction of
supersymmetric gauge-Higgs unification is a specific boundary
condition on the Higgs mass parameters at the compactification scale.
In fact, the predicted boundary condition is same as
the one which has been obtained in supersymmetric
pseudo-Goldstone Higgs boson models \cite{pNG}.
This can be easily understood by noting that higher-dimensional
gauge symmetry constrains the K\"ahler potential
of Higgs superfields in the same way as non-linear
global symmetry in pseudo-Goldstone Higgs boson models does.

Although the boundary condition on the Higgs mass parameters
predicted by gauge-Higgs unification
is independent of the details of
supersymmetry (SUSY) breaking, its phenomenological viability
severely depends on SUSY breaking mechanism since
the Higgs mass parameters at the weak scale
receive large radiative corrections
depending on other sparticle masses \cite{higgs}.
It turns out that  the condition of electroweak symmetry
breaking severely restricts the
parameter space of the model for SUSY breaking
scenarios which can be realized naturally in
orbifold field theories.
If $N=1$ SUSY breaking masses are much smaller
than the compactification scale, which will be assumed
throughout this paper, the $N=1$ SUSY breaking
can be described by the auxiliary components of
4D $N=1$ superfields. Then there can be
two distinctive sources of SUSY breaking for visible fields,
one mediated by the zero modes of bulk superfields propagating in
5D spacetime and the other mediated by 4D brane superfields
confined on the orbifold fixed points.
An attractive possibility is that SUSY breaking is
mediated dominantly by the bulk radion superfield $T$, which
is equivalent to the Scherk-Schwarz (SS) SUSY breaking by boundary
conditions \cite{SS,SS2}.
Another source of SUSY breaking in bulk is
 the auxiliary component of 4D supergravity (SUGRA) multiplet
which can be parameterized by the $F$-component
of the chiral conformal-compensator superfield $\Omega$
\cite{compensator}.
Classical conformal invariance ensures that
soft scalar masses, gaugino masses and trilinear $A$-parameters
are not affected by $F^\Omega$ at tree level,
however the Higgs $\mu$ and $B\mu$ parameters
receive contributions from
$F^\Omega$ even at tree level \cite{anomaly}.
As we will see, if the 4D effective theory of the model is
given by the minimal supersymmetric standard model (MSSM)
with soft SUSY breaking parameters induced by the $F$-components
of  the radion superfield $T$ and the compensator superfield
$\Omega$,
a correct electroweak symmetry breaking can {\it not} be achieved
for  reasonable range of parameters of the model.
Therefore one needs an additional SUSY breaking
other than those from the SS mechanism and
the 4D SUGRA multiplet, e.g. the SUSY breaking
mediated by a brane superfield,
in order to achieve electroweak symmetry breaking in 5D gauge-Higgs
unification models whose 4D effective theory corresponds to
the MSSM.
When there exists a supersymmetry breaking mediated by
brane superfield, sizable portion of the parameter space can give
a correct electroweak symmetry breaking.

This paper is organized as follows. In Section II, we discuss
SUSY breaking masses in 5D gauge-Higgs unification models,
particularly the Higgs mass parameters, in the framework of
4D effective action in $N=1$ superspace. In Section III, we
perform a numerical analysis for electroweak symmetry breaking
at the weak scale under the assumption that
the 4D effective theory below the compactification scale is given by the
MSSM.
Section IV is the conclusion.

\section{susy breaking masses in 5d gauge-higgs unification  models}

The most efficient way to compute SUSY breaking masses is to derive
the 4D effective action in $N=1$ superspace which contains the
SUSY-breaking messenger superfields explicitly
\cite{compensator}.
In this section, we derive the (part of) 4D effective action
of a 5D theory compactified on $S^1/Z_2\times Z_2^\prime$
in which the 4D Higgs fields
originate from 5D vector supermultiplets, and discuss SUSY breaking
masses induced by the auxiliary components of the 4D SUGRA multiplet,
the radion superfield, and
also generic chiral brane superfields.

In $N=1$ superspace, a 5D vector multiplet is represented by a
vector superfield $V$ and a chiral superfield $\Sigma$.
The orbifold boundary conditions of
these $N=1$ superfields are given by
\bea
&& V^a(-y)=z_aV^a(y)\,, \quad V^a(-y')=z'_aV^a(y')\,,
\nonumber \\
&& \Sigma^a(-y)=-z_a\Sigma^a(y)\,,\quad \Sigma^a(-y')=
-z'_a\Sigma^a(y')\,,
\eea
where $y'=y-\pi$, $z_a=\pm 1$ and $z_a'=\pm 1$.
In gauge-Higgs unification models, 4D gauge bosons originate from
$V^a$ with $z_a=z'_a=1$,
while 4D Higgs bosons originate from
$\Sigma^a$  with $z_a=z'_a=-1$.
The 5D gauge transformations associated with $\{V^a,\Sigma^a\}$
are given by
\beq
\label{gtransformation}
e^{V}\,\rightarrow\, e^\Lambda e^V e^{\Lambda^\dagger}\,,
\quad
\Sigma\,\rightarrow\, e^\Lambda \left(\Sigma
-\sqrt{2}\partial_y\right)e^{-\Lambda}\,,
\eeq
where $V=V^aT^a,\Sigma=\Sigma^aT^a$,
and $\Lambda=\Lambda^aT^a$ denotes the chiral gauge transformation
superfield satisfying the orbifold boundary condition:
$$
\Lambda^a(-y)=z_a\Lambda^a(y)\,,
\quad
\Lambda^a(-y')=z'_a\Lambda^a(y')\,.
$$
%Here we use the convention in which
%the fundamental domain of $S^1/Z_2\times Z_2'$ is
%given by $0\leq y\leq \pi$.

The 5D action of $V^a$ and $\Sigma^a$ in $N=1$ superspace is given by
\cite{marti}
\bea
\label{bulkaction}
S_{\rm bulk}=\int d^5x \,\left[\, \int d^4 \theta
\,\frac{1}{g_{5}^2}\frac{1}{T+T^*}\left(
\, \partial_y V^a-\frac{1}{\sqrt{2}}(\Sigma^a+\Sigma^{a *})\,
\right)^2
+\int d^2\theta \,
\frac{T}{4g_{5}^2}W^{a\alpha}W^a_\alpha
+...\,\,\right]\,,
\eea
where $g_5^2$ is the (unified) 5D gauge coupling with mass dimension $-1$
and the radion superfield $T$ is given by
$$
T=R+iB_5+\theta \Psi^2_5+\theta^2 F^T\,,
$$
where $R$ is the orbifold radius,
$B_5$ and $\Psi^2_5$ are the fifth-components of the graviphoton
$B_M$ and the symplectic Majorana gravitini $\Psi^i_M$ ($i=1,2$).
Here we limit ourselves to the terms which are bilinear
in $V^a$ and $\Sigma^a$.
In addition to the above bulk action,
there can be interactions of
$V^a$ and $\Sigma^a$ at the orbifold fixed points,
particularly the interactions
with chiral brane superfields which have
nonzero SUSY breaking auxiliary components.
Those fixed-point interactions are restricted
also by the 5D gauge symmetry (\ref{gtransformation})
and 5D general covariance, and generically given by
\bea
\label{braneaction2}
S_{\rm brane}&=& \int d^5x \,\left[\,
\int d^4\theta \,
\frac{\delta(y)\Delta Y_a({\cal Z},{\cal Z}^*)
+\delta(y-\pi)\Delta Y'_a({\cal Z}',{\cal Z}^{\prime *})}{
g_5^2(T+T^\dagger)^2}
\left(\p_y V^a-\frac{1}{\sqrt{2}}(\Sigma_a+\Sigma_a^*)\right)^2
\right.\nonumber \\
&& \left.\quad\quad \quad
+\int d^2\theta \,\frac{1}{4g_5^2}\left[\,
\delta(y)\omega_a({\cal Z})+\delta(y-\pi)\omega'_a({\cal Z}')\,\right]
W^{a\alpha}W^a_\alpha+...\,\right]
\eea
where ${\cal Z}$ and ${\cal Z}'$ stand for
generic SUSY breaking brane superfields
at $y=0$ and $y=\pi$, respectively.

The 4D effective action
of the gauge and Higgs zero modes at the compactification scale
can be written as
\bea
\label{4daction1}
S_{\rm eff}&=&
\int d^4x \left[\,\int d^4\theta
\,e^{\Omega+\Omega^\dagger}\left(\,
Y_{H_1}H_1^\dagger H_1+
Y_{H_2}H_2^\dagger H_2+
\gamma_HH_1H_2+\gamma^*_HH^\dagger_1H^\dagger_2
\,\right)\right.
\nonumber \\
&&\quad\quad\quad+\left.\int d^2\theta \,\frac{1}{4}f_aW^{a\alpha}W^a_\alpha
+...\,\right]
\eea
where the gauge kinetic function $f_a$
is a holomorphic function of $\{{\cal Z}_A\}=
\{T, {\cal Z},{\cal Z}'\}$, while the Higgs wavefunction
coefficients $Y_{H_i}$ and $\gamma_H$
depend on both $\{{\cal Z}_A\}$ and $\{{\cal Z}_A^*\}$.
Here $H_1$ and $H_2$
are the two MSSM Higgs superfields originating from
\beq
\Sigma^1=-(H_1+H_2)\,,\quad
\Sigma^2=-i(H_1-H_2)\,,
\eeq
with $z_{1,2}=z'_{1,2}=-1$, while $W^{a\alpha}$ are the chiral gauge
superfields for the zero modes of $V^a$ with $z_a=z_a=1$.
Note that  we have introduced the chiral conformal-compensator
superfield $e^\Omega$ to parameterize the
SUSY breaking by the auxiliary component of 4D SUGRA multiplet
\cite{compensator}.
It is then straightforward to find
\bea
\label{4daction2}
&& Y_{H_1}=Y_{H_2}=\gamma_H=\frac{2\pi }{g_{5}^2(T+T^*)}
\left[\,1+
\frac{\Delta Y_H({\cal Z}, {\cal Z}',
{\cal Z}^*,{\cal Z}^{\prime *})}{\pi(T+T^*)}\,\right]\,,
\nonumber \\
&& f_a=\frac{1}{g_{5}^2}\,\left[
\,\pi T+\omega_a({\cal Z})+\omega'_a({\cal Z}')\,\right]\,,
\eea
where $\Delta Y_H$ represents the contributions
from $\Delta Y_a$ and $\Delta Y'_a$ in (\ref{braneaction2}) for
$\Sigma^a$ from which the Higgs fields originate.

A simple dimensional analysis suggests that
the vacuum expectation values of $\omega_a, \omega_a',
\Delta Y_a$ and $\Delta Y_a'$ are generically of
the order of the cutoff length scale,
and thus
\bea
\label{nda1}
\frac{\Delta Y_H({\cal Z},{\cal Z}',{\cal Z}^*,{\cal Z}^{'*})}
{\pi (T+T^*)}&=&{\cal O}(1/\pi R\Lambda)\,,
\nonumber \\
\frac{\omega_a({\cal Z})+\omega'_a({\cal Z}')}{\pi T} &=
&{\cal O}(1/\pi R\Lambda)\,,
\eea
where $\Lambda$ denotes the cutoff mass scale.
By construction, $\pi R\Lambda$ should be
bigger than the unity, however its precise value
depends on the radion stabilization
mechanism in the theory.
Throughout this paper, we will assume that the theory is strongly
coupled at $\Lambda$ \cite{nda}, and also
\beq
\label{nda2}
\pi R\Lambda={\cal O}(8\pi^2)\,,
\eeq
which are consistent with each other.
Under these assumptions, the contributions from brane actions
to the SUSY-preserving components of $Y_H$ and $f_a$
are suppressed by ${\cal O}(1/8\pi^2)$
compared to the bulk contributions.
Note that this does {\it not} mean that SUSY-breaking masses are
dominated also by the bulk contributions since
the $F$-components of ${\cal Z}$ and/or ${\cal Z'}$
can be significantly bigger
than the $F$-components of $T$ and $\Omega$.

One important consequence of gauge-Higgs unification is that
\bea
\label{prediction}
Y_{H_1}=Y_{H_2}=\gamma_H\,.
\eea
Obvioulsy the contributions to $Y_{H_{1,2}}$ and $\gamma_H$
from the bulk action (\ref{bulkaction})
obey the above relation due to the 5D gauge symmetry
(\ref{gtransformation}) unifying the Higgs fields with gauge fields.
Generically this bulk gauge symmetry is broken down to
a subgroup  at the orbifold fixed point by boundary conditions,
so one may expect the relation (\ref{prediction})
is broken  by the contributions from the brane action
(\ref{braneaction2}).
However still
(\ref{braneaction2}) is constrained by (\ref{gtransformation}),
in particular, by the non-linear gauge symmetry under which
\bea
\label{nonlinear}
&&\Sigma^{1}\quad \rightarrow \quad \Sigma^1+\sqrt{2}\p_y\Lambda^1+...\,,
\nonumber \\
&&\Sigma^{2}\quad \rightarrow \quad \Sigma^2+\sqrt{2}\p_y\Lambda^2+...
\,,\nonumber \\
&& V^{1,2}\quad \rightarrow\quad
 V^{1,2}+\Lambda^{1,2}+(\Lambda^{1,2})^\dagger+...\,,
\eea
where $\Lambda^{1,2}(-y)=-\Lambda^{1,2}(y)$ and
$\Lambda^{1,2}(-y')=-\Lambda^{1,2}(y')$  are parity-odd gauge transformation
superfields, and the ellipses denote the transformations not relevant
for constraining the terms which are bilinear in the Higgs superfields.
This non-linear gauge symmetry constrains the brane actions
at the fixed points in such a way that
the contributions to $Y_{H_{1,2}}$ and $\gamma_H$
from the brane actions
satisfy the relation (\ref{prediction}).
Note that the non-linear gauge symmetry (\ref{nonlinear})
is realized for 5D fields, so
{\it not} valid if the massive Kaluza-Klein modes are integrated
out. As a result, the relation (\ref{prediction})
is valid {\it only} at the compactification scale,
but is modified by radiative corrections at lower energy scales.

Let us discuss SUSY breaking scenarios possible in 5D orbifold field theories.
Since we are assuming that SUSY breaking scale is much lower
than the compactification scale, any SUSY breaking can be
described by the auxiliary components of
4D messenger superfields.
In 5D models under consideration,
possible messenger superfields include
the 4D SUGRA multiplet, the radion superfield $T$ and
also some set of brane superfields $\{{\cal Z},{\cal Z}'\}$.
In the compensator formulation of 4D SUGRA,
SUSY breaking by the  4D SUGRA
multiplet can be described by the $F$-component of
the chiral conformal-compensator superfield \cite{compensator}
 \beq
e^\Omega
=e^{\Omega_0}(1+\theta^2 F^\Omega).
\eeq
In the Einstein frame, $\Omega_0={K/6}$ where
$K$ is the K\"ahler potential.
Then the compensator $F$-component is given by
\bea
F^\Omega=m_{3/2}^*+\frac{1}{3}\frac{\partial K}{\partial
{\cal Z}_A}F^A\,,
\eea
where $m_{3/2}=e^{K/2}W$ is the gravitino mass
for the superpotential $W$ and
$F^A$ is the $F$-component of $\{{\cal Z}_A\}=\{T, {\cal Z},
{\cal Z}'\}$. Once the K\"ahler potential $K$ and
the superpotential $W$ of 4D effective theory are known,
$F^A$ in the Einstein frame is determined
to be \cite{munoz}
\bea
F^A=-e^{K/2}K^{AB}
\left(\frac{\partial W}{\partial {\cal Z}_B}+
\frac{\partial K}{\partial {\cal Z}_B}W\right)^*\,,
\eea
where $K^{AB}$ is the inverse of the K\"ahler metric
$K_{AB}=(\partial^2 K/\partial {\cal Z}_A
\partial {\cal Z}^*_B)$.

The SUSY breaking mediated by $F^T$ corresponds to the Sherk-Schwarz
SUSY breaking by twisted boundary condition \cite{SS2}, thus is a natural
candidate for SUSY breaking in  models with extra dimension.
For 4D fields
originating from 5D bulk fields, all
soft SUSY breaking masses generically receive a contribution
of  ${\cal O}(F^T/R)$ at tree approximation.
As for the SUSY breaking mediated by $F^\Omega$,
{\it only} the Higgsino mass $\mu$ and the bilinear Higgs
coefficient $B$ can receive a contribution of the order of
$F^\Omega$ at tree level \cite{munoz},
while the gaugino masses
$M_a$, soft scalar masses $m_\phi$, and trilinear
scalar coefficients $A$ get the conformal anomaly-mediated contributions
of ${\cal O}(F^\Omega/8\pi^2)$
\cite{anomaly}.
The anomaly-mediated scenario corresponds to
the case that $F^\Omega/8\pi^2\gg F^T/R$, so
$M_a$, $m_\phi$ and $A$ are dominated by
the conformal anomaly-mediated contributions.
However the anomaly-mediated scenario is not
acceptable in gauge-Higgs unification model since $\mu$ and $B$ always
receive a contribution of ${\cal O}(F^\Omega)$, thus
become ${\cal O}(8\pi^2M_a)$
unless an unnatural cancellation is assumed, which
would be too large to allow
a correct electroweak symmetry breaking.
In this paper, we  will  assume
that $F^\Omega$ is of the order of $F^T/R$ or less, and
thus ignore the conformal anomaly-mediated
contributions of
${\cal O}(F^\Omega/8\pi^2)$.

%The ratio $F^\Omega/F^T$ depends on the details of
%the radion stabilization mechanism.
%In this paper, we are not attempting
%to compute $F^C/F^T$, so treat it as a free parameter.

Following the standard method to compute soft SUSY breaking masses
in 4D SUGRA \cite{munoz}, we find that the gaugino and
Higgs mass parameters
for generic $f_a$ and $Y_H$ are given by
\bea
\label{eq:tree_potential}
{\cal L}_{\rm soft}&=&
-(\,\frac{1}{2}M_a\lambda_a\lambda_a+\mu\tilde{H}_1\tilde{H}_2
+{\rm h.c}\,)\nonumber\\
&&-(|\mu|^2+m_{H_1}^2)|H_1|^2-
(|\mu|^2+m_{H_2}^2)|H_2|^2+\left(\,
B\mu H_1H_2+{\rm h.c}\,\right)\,
\eea
where
\bea
M_a&=& -\frac{1}{2 {\rm Re}(f_a)}F^A\frac{\partial f_a}{\p {\cal Z}_A}
\,,
\nonumber \\
\mu^*&=& -F^\Omega - F^A\f{\p \ln Y_H}{\p {\cal Z}_A}\,,
\nonumber \\
m_{H_1}^2+|\mu|^2&=&
m_{H_2}^2+|\mu|^2 = -B\mu=
|\mu|^2- F^AF^{B*}\f{\partial^2 \ln Y_H}{\p {\cal Z}_A
\p {\cal Z}_B^*}\,,
%\nonumber \\
%B\mu&=&\left|F^\Omega +F^A\f{\p \ln Y_H}{\p {\cal Z}_A}\right|^2
%-F^AF^{B*}\f{\partial^2 \ln Y_H}{\p {\cal Z}_A
%\p {\cal Z}^*_B}\,,
\eea
for {\it canonically normalized} gauginos $\lambda_a$, Higgs bosons $H_1$
and $H_2$, and Higgsinos $\tilde{H}_1$ and $\tilde{H}_2$.
Note that the relation between Higgs mass parameters
%\footnote{
%In this paper, we use the field basis for which
%the Higgs mass relation from gauge-Higgs unification
%is given by (\ref{prediction1}), not by the one obtained by
%replacing $-B\mu$ in (\ref{prediction1}) by $B\mu$.}
\bea
\label{prediction1}
m^2_{H_1}+|\mu|^2=m_{H_2}^2+|\mu|^2=-B\mu
\eea
is a consequence of (\ref{prediction}),
thus can be considered as a prediction of
gauge-Higgs unification
which is independent of the details of SUSY breaking mechanism.
As we have noticed, (\ref{prediction}) is a consequence of
the non-linear 5D gauge symmetry (\ref{nonlinear}) which is valid
only at scales above the compactification scale, so
the above relation
corresponds to a boundary condition at the compactification scale.
In fact, the same relation between Higgs mass parameters
has been obtained before in the context of
supersymmetric pseudo-Goldstone Higgs models.
This is not surprising because the non-linear 5D gauge symmetry
in gauge-Higgs unification models plays the same role as the
non-linear global symmetry in pseudo-Goldstone Higgs models.

One attractive way to break SUSY in orbifold field theory
is the Scherk-Schwarz mechanism to impose different boundary
conditions for different fields in the same supermultiplet
\cite{SS}.
It has been pointed out that the SS  breaking
is equivalent to the SUSY breaking
by $F^T$ \cite{SS2}.
Here we consider a more generic situation that there exist
additional contributions from $F^\Omega$ to the Higgs mass parameters,
which are generically of ${\cal O}(F^T/R)$.
In this SUSY breaking scenario mediated by
$F^T$ and $F^\Omega$, the gaugino masses and Higgs mass parameters
are given by
\bea
\label{higgs}
M_a&=& -\frac{F^T}{2R}\,,
\nonumber \\
\mu^*&=& -F^\Omega+\frac{F^T}{2R}\,,
\nonumber \\
m_{H_1}^2+|\mu|^2&=&m_{H_2}^2+|\mu|^2=-B\mu =
-|M_a|^2+|\mu|^2\,,
\eea
where we have ignored the contributions from
brane actions under the assumption of
(\ref{nda1}) and (\ref{nda2}), and also
the loop-suppressed anomaly-mediated contributions of
${\cal O}(F^\Omega/8\pi^2)$.

To study the electroweak symmetry breaking,
%for the Higgs mass parameters satisfying (\ref{prediction1})
%at the compactification scale,
one needs also information
on the sfermion masses and trilinear $A$-parameters
for the quark and lepton superfields having large
Yukawa coupling.
In fact, the sfermion masses and $A$-parameters
(at the compactification scale) in gauge-Higgs unification model
depend highly on the details of the mechanism
to generate hierarchical Yukawa couplings
and flavor mixings, i.e.
on the details of  the quasi-localization of bulk fermions
and the mixings with brane fermions.
However, to have a large top-quark Yukawa coupling
$y_t\approx g_{_{GUT}}$, there should not be
any sizable suppression of the top-quark Yukawa coupling
by quasi-localization
and/or the mixing with brane fermions.
In this case, the K\"ahler metrics of
the $SU(2)$-doublet top quark superfield $Q_3$ and
the $SU(2)$-singlet top quark superfield $U_3$
are given by \cite{future}
\bea
Y_{Q_3} &\approx& \frac{\pi}{2} (T+T^*) +\Delta Y_{Q_3}({\cal Z},{\cal Z}',
{\cal Z}^*, {\cal Z}^{\prime *})
\nonumber \\
Y_{U_3}&\approx & \frac{\pi}{2} (T+T^*)+\Delta Y_{U_3}({\cal Z},{\cal Z}',
{\cal Z}^*,{\cal Z}^{\prime *})
\eea
where $\Delta Y$ represent the contributions from brane actions.
The resulting soft stop masses and $A$-parameter are given by
\bea
{\cal L}_{\rm soft}=
-m^2_{Q_3}|\tilde{Q}_3|^2-m_{U_3}^2|\tilde{U}_3|^2+
(\,A_ty_tH_2\tilde{Q}_3\tilde{U}_3 +{\rm h.c}\,)
\eea
 where
\bea
m^2_{Q_3}&=&-F^AF^{B*}\frac{\partial^2 \ln Y_{Q_3}}{\partial
{\cal Z}_A{\cal Z}^*_B}\,,
\nonumber \\
m^2_{U_3}&=&-F^AF^{B*}\frac{\partial^2\ln Y_{U_3}}{
\partial {\cal Z}_A{\cal Z}^*_B}\,, \nonumber \\
A_t&=& -F^A\frac{\partial\ln (Y_HY_{Q_3}Y_{U_3})}{\partial{\cal Z}_A}\,.
\eea
When SUSY breaking is dominated by
$F^T$ and $F^\Omega$, we have
\bea
\label{stop}
A_{t}&\approx& \frac{F^T}{2R}=-M_a\,,
\nonumber \\
m^2_{{Q}_3}&\approx&m^2_{{U}_3}\approx
\left|\frac{F^T}{2R}\right|^2\,.
\eea
If $\tan\beta=\langle H^0_2\rangle/\langle H^0_1\rangle$ is large,
the $b$-quark and $\tau$-lepton also have
a large Yukawa coupling, and then
 the sbottom and stau
have similar soft masses
and $A$-parameters \cite{future}.

\medskip

\section{analysis of electroweak symmetry breaking}

\begin{figure}[t]
\begin{center}
\begin{minipage}{16cm}
\centerline{
{\hspace*{-.2cm}\psfig{figure=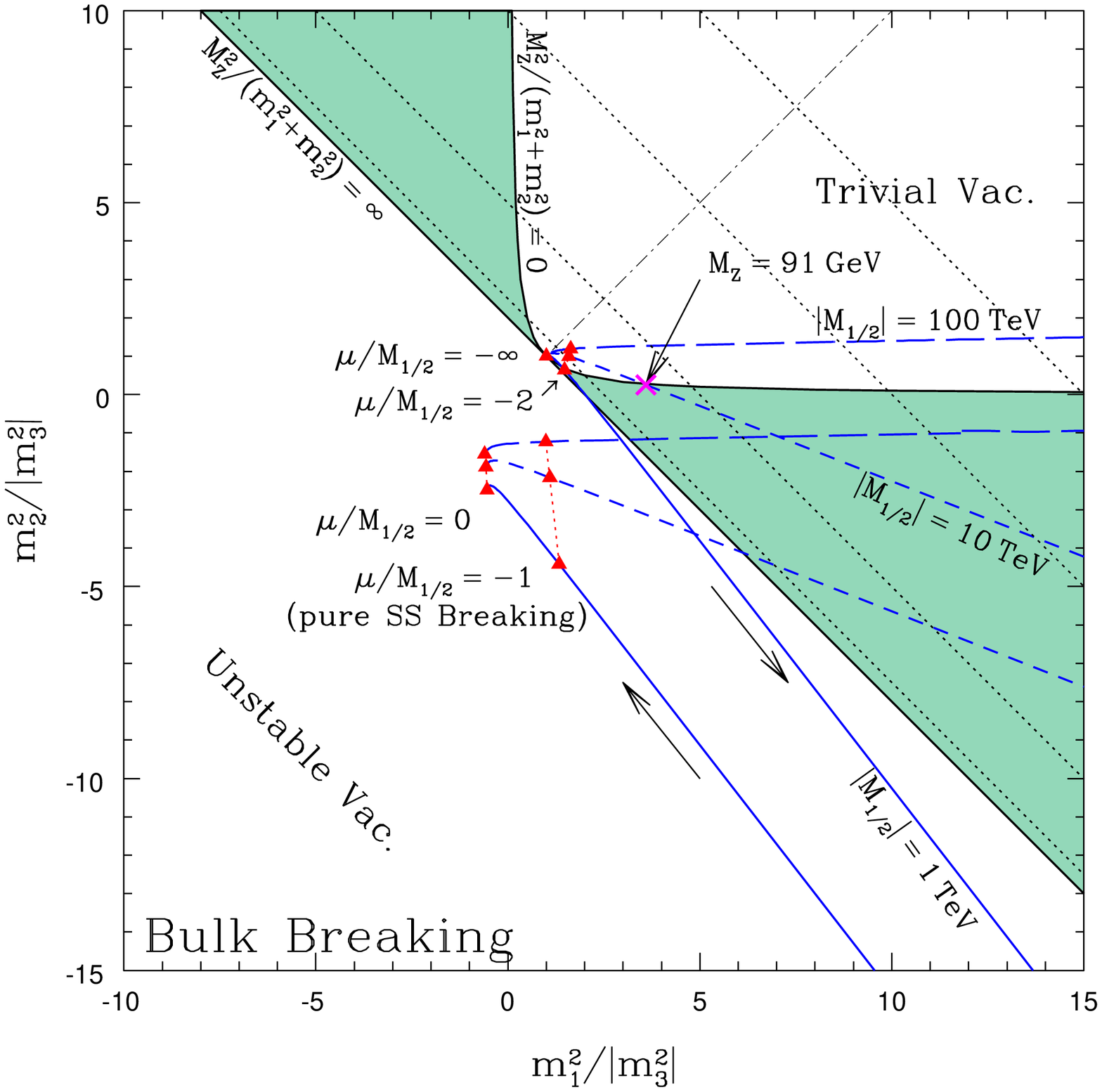,angle=0,width=8.0cm}}
{\hspace*{-.2cm}\psfig{figure=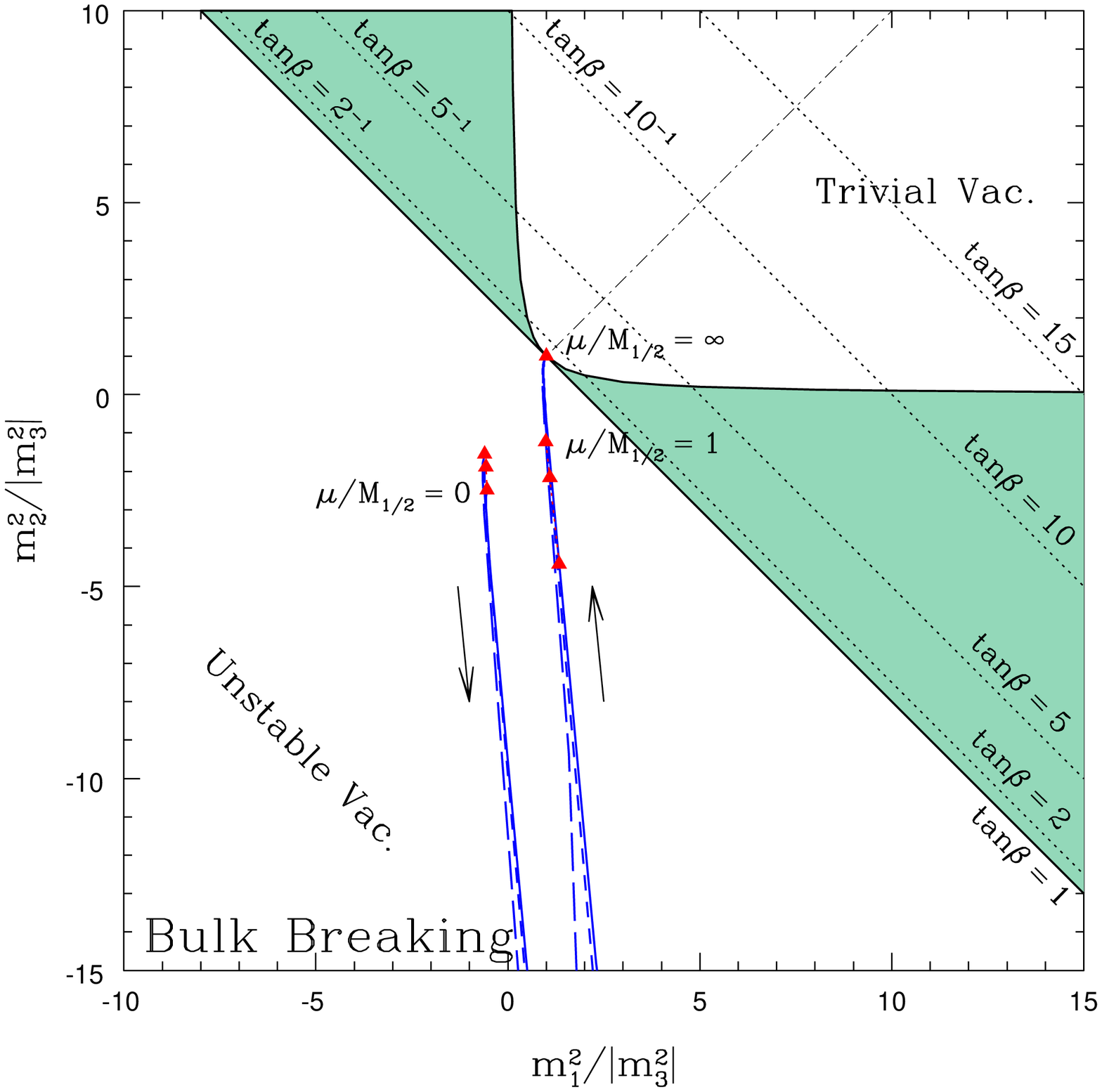,angle=0,width=8.0cm}}
}
\caption{
Results of the analysis of the RG-improved tree level Higgs
potential for the bulk SUSY breaking mediated by
$F^T$ and $F^\Omega$.
Here $m_i^2=m_{H_i}^2+|\mu|^2$ ($i=1,2$)
and $m_3^2=B\mu$ at $M_{\rm SB}$.
Electroweak symmetry is broken in the shaded regions.
The solid, dashed and long-dashed lines indicate the
solutions of RG equations for $|M_{1/2}|=1, \,10,\, 100$ TeV, respectively.
The parameter ratio $\mu/M_{1/2}$ is varying from
$-\infty$ to $0$ in Fig.1a (left),
and from $0$ to $\infty$ in Fig.1b (right).
Note that the pure Sherk-Schwarz SUSY breaking scenario, i.e.
$F^\Omega=0$, gives $\mu/M_{1/2}=-1$, which is clearly outside the symmetry
breaking shaded region.
The long arrows show the direction to which $\mu/M_{1/2}$ increases.
The stop soft parameters at $M_X$ are chosen to be
$A_t/M_{1/2}=-1$ and $m^2_{Q_3}=m^2_{U_3}=M_{1/2}^2$
as given in (\ref{stop}), and the top quark Yukawa coupling
$y_t(m_t)=0.98$.
\label{fig:tree_rewsb}}
\end{minipage}
\end{center}
\end{figure}

Given the predictions for soft SUSY breaking masses at the
compactification scale, we analyze electroweak symmetry breaking
 by numerically solving the renormalization group (RG) equations
 down to the symmetry breaking scale, $M_{\rm SB}$,
 and minimizing the effective Higgs potential.
We identify the compactification scale as the gauge unification scale,
$M_X = 2\times 10^{16}$ GeV together with an assumption that
GUT scale gaugino masses are universal, $M_a(M_X)=
M_{1/2}$, and  $M_{\rm SB}$ as $\sqrt{m_{\tilde{t}_1} m_{\tilde{t}_2}}$
following the standard procedure where $m_{\tilde{t}_{1,2}}$
are the stop mass eigenvalues\cite{Baer:1997yi}.
It is further assumed that the effective theory between $M_X$ and $M_{\rm SB}$
is given by the MSSM which is reduced to
the standard model at scales below $M_{\rm SB}$.
We will use the parameter convention in which
the gauge-Higgs unification prediction for Higgs mass parameters
takes the form (\ref{prediction1}).

To obtain the gauge and Yukawa couplings at $M_X$,
we solve two-loop RG running \cite{2loop_rge}
 from the electroweak scale to $M_X$, however
the matching at $M_{\rm SB}$
 is performed without including the superparticle threshold corrections.
Using the boundary conditions at $M_X$,
 the Higgs mass parameters at $M_{\rm SB}$
 is calculated using one-loop RG equations for dimensionful parameters
  and two-loop RG equations for the gauge and Yukawa couplings.
Given the inputs of dimensionless parameters at
the electroweak scale and dimensionful
parameters at $M_X$, a self-consistent value of $M_{\rm SB}$
is extracted by numerical iteration starting from $M_{\rm SB}$
taken to be the top quark mass.

A simple way to analyze the electroweak symmetry breaking
is to compute the RG-improved tree-level Higgs potential
at the electroweak scale which can be written as
\beq
V=m_1^2|H_1^0|^2+m_2^2|H_2^0|^2-\l(
B\mu H_1^0H_2^0+\mbox{c.c.}\r)+
\frac{1}{8}(g_1^2+g_2^2)\l(|H_1^0|^2-|H_2^0|^2\r)^2\,,
\eeq
where $m_{1,2}^2=m_{H_{1,2}}^2+|\mu|^2$.
To develop non-trivial vacuum ($\langle H^0_{1,2}\rangle \neq 0$)
 and stabilize the flat direction of the D-term potential,
the Higgs mass parameters should satisfy
\bea
\label{eq:rewsb_condition}
m_1^2 m_2^2 - |B\mu|^2 < 0 &,&~~~
m_1^2 + m_2^2 - 2 |B\mu| > 0.
\eea
Under this conditions, $M_Z$ and $\tan\beta$ at the minimum of the potential
acquire the following values:
\bea
\label{eq:rewsb_relation}
M_Z^2 = (m^2_1+m^2_2)\left[\left\{1+4\left(\frac{m^2_1 m^2_2 - |B\mu|^2}{(m^2_1-m^2_2)^2}\right)\right\}^{-\frac{1}{2}}-1\right]&,&
\tan\beta = \frac{m^2_1+m^2_2}{|B\mu|} - \frac{1}{\tan\beta}.
\eea
This parameter region for stable electroweak symmetry
breaking is given by
the shaded area in Fig.\ref{fig:tree_rewsb}.

Let us now consider the SUSY breaking  by
$F^T$ and $F^\Omega$, i.e. the Sherk-Schwarz SUSY breaking with
additional contributions to Higgs mass parameters
from the auxiliary component of the 4D SUGRA multiplet.
In this scenario,  SUSY breaking parameters at the
GUT scale  are given by
(\ref{higgs}) and (\ref{stop}).
The resulting solutions of RG equations for
$|M_{1/2}|=$ 1 TeV, 10 TeV and 100 TeV
are depicted in Fig.\ref{fig:tree_rewsb} by
the solid, dashed and long-dashed lines, respectively.
For numerical analysis, we use
the top quark Yukawa coupling $y_t(m_t)=0.98$,
and also the GUT scale predictions
$A_t/M_{1/2}=-1$ and $m_{Q_3}^2=m_{U_3}^2=M_{1/2}^2$.
For simplicity, we ignored the effects of the $b$-quark and $\tau$-lepton
Yukawa couplings, so the results can be somewhat
modified for large $\tan\beta$.
In Fig.\ref{fig:tree_rewsb}a (left),
$\mu(M_X)/M_{1/2}$ varies from $-\infty$ to 0,
while it varies from 0 to $\infty$
in Fig.\ref{fig:tree_rewsb}b (right).
For each value of $|M_{1/2}|$, we have two lines
since  $B\mu (M_{\rm SB})$ crosses zero,
thereby the solution goes to infinity
at some value of $\mu(M_X)/M_{1/2}$.
As $|M_{1/2}|$ is decreasing
from 100 TeV to 1 TeV, the solutions sweep the shaded region
of Fig.\ref{fig:tree_rewsb}a from upper right to lower right,
while there exists no solution in the shaded region
of Fig.\ref{fig:tree_rewsb}b.
This   indicates that
 $1\,\,{\rm TeV} \ler |M_{1/2}| \ler 100\,\,{\rm TeV}$ with $\mu/M_{1/2}<0$
is required to obtain a symmetry breaking vacuum.
Still to get the correct value of
$M_Z$, one needs a fine-tuned value of
$\mu(M_X)$ for each value of $|M_{1/2}|$.

The pure Sherk-Schwarz SUSY breaking scenario gives
(\ref{higgs}) and (\ref{stop}) with $F^\Omega=0$,  so
 $\mu(M_X)=-M_{1/2}$.
As can be seen in Fig.\ref{fig:tree_rewsb}a, such parameter value
develops an unstable vacuum, and thus should be excluded.
With $F^\Omega\neq 0$, $\mu/M_{1/2}$ can have an arbitrary
value in principle.
However for the case with $\mu/M_{1/2}>0$,
as can be seen in Fig.\ref{fig:tree_rewsb}b,
the whole parameter space of (\ref{higgs}) and (\ref{stop})
develops an unstable vacuum, thus should be excluded also.
A stable vacuum with correct value of $M_Z$
can be obtained for $-2 \lesssim \mu(M_X)/M_{1/2} <-1$,
but only for an abnormally large gaugino mass
$$
|M_{1/2}|\gtrsim 10 \,\,\mbox{TeV}\,.
$$
In this case, we need an unnatural fine-tuning of $\mu(M_X)$
with an accuracy $\delta\mu/\mu\lesssim 10^{-4}$
in order to get $M_Z=91$ GeV, which is
hard to be accepted.
We thus conclude that
a correct electroweak symmetry breaking can not be achieved
in SUSY breaking scenarios mediated by $F^T$ and $F^\Omega$ alone,

% which requires a compensator contribution to ,
% $0 > \mu/M_a+1 = F^{\Omega}(\frac{F^T}{2R})^{-1} \gtr -1$
% fine-tuned for $|M_a|=|\frac{F^T}{2R}| \gtr 10\,{\rm TeV}$
%\footnote{It is interesting to observe that $m^2_2$ crosses zero almost
% simultaneously with $B\mu$ for $M_a\simeq 100$ TeV and $\mu(M_X)
% \simeq -1.2 M_a$. In this case, hierarchically small $M_Z$ is naturally
% realized as $M_Z^2 \simeq 2 m^2_1 (\frac{|B\mu|}{m^2_1})^2$.
% However, $\tan\beta\simeq \frac{m^2_1}{|B\mu|} \sim 10^3$ is too large to
% reproduce a realistic value of $M_b$.}.
%

%
\begin{figure}[t]
\begin{center}
\begin{minipage}{12cm}
\centerline{
{\hspace*{-.2cm}\psfig{figure=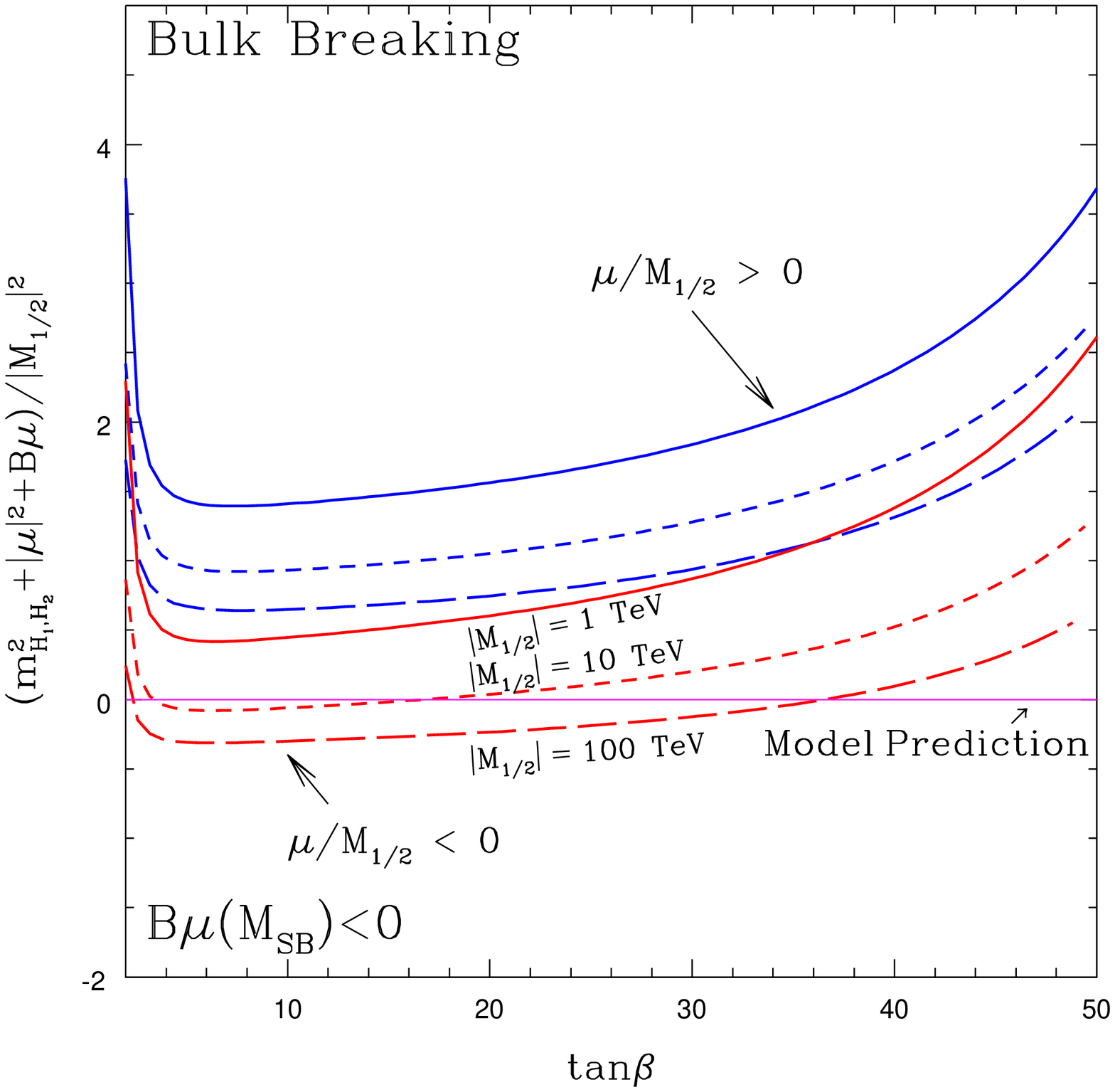,angle=0,width=8.0cm}}
{\hspace*{-.2cm}\psfig{figure=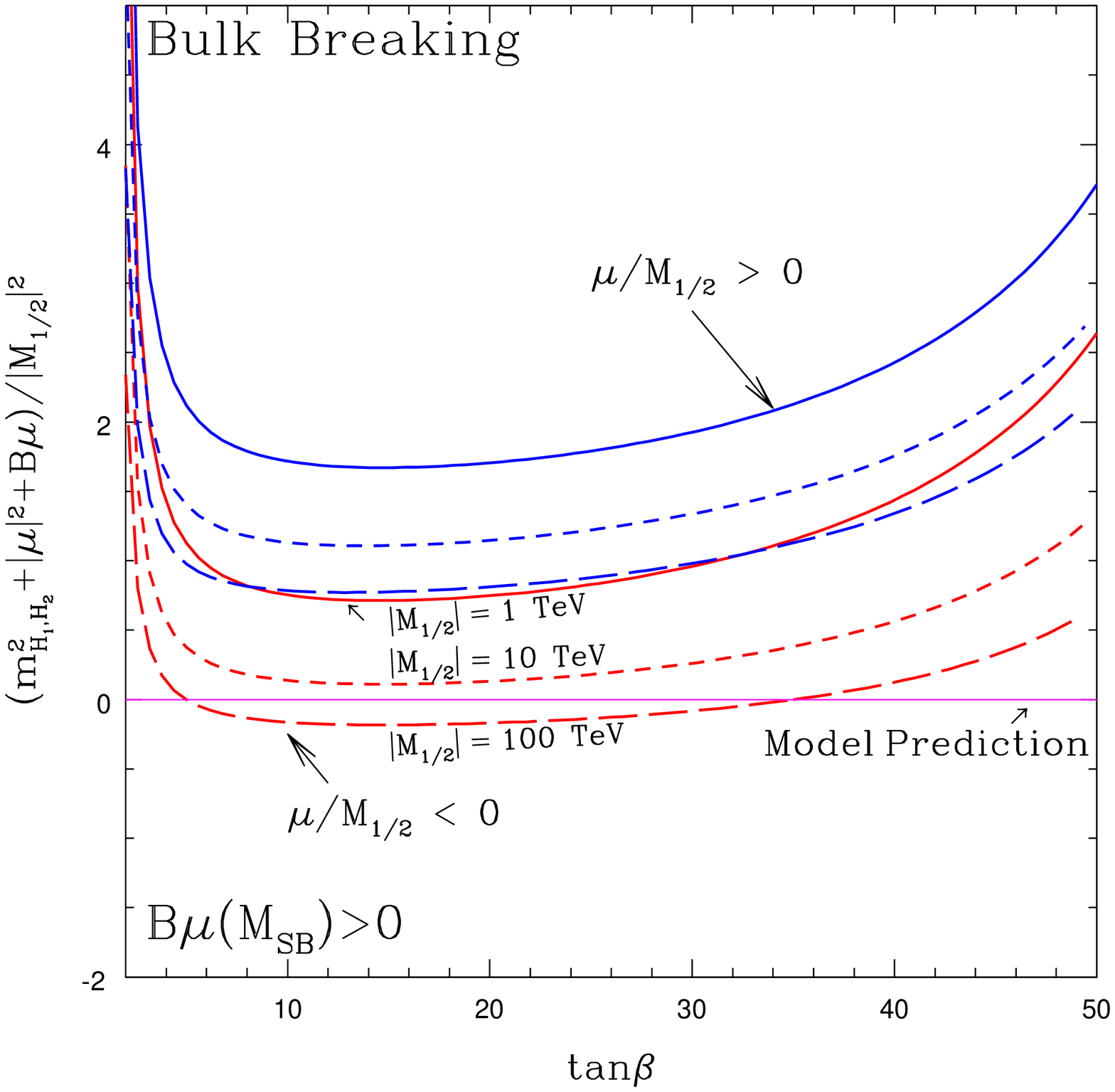,angle=0,width=8.0cm}}
}
\caption{Results of the alternative analysis for the bulk
SUSY breaking mediated by $F^T$ and $F^\Omega$.
The value of $(m_{H_1, H_2}^2+|\mu|^2+B\mu)/|M_{1/2}|^2$ at $M_X$ is
obtained for $\mu$ and $B$ at $M_{\rm SB}$ giving $M_Z=91$ GeV, and is
is depicted as a function of $\tan\beta$.
The solid, dashed and long-dashed curves represent the results
for $|M_{1/2}|= 1,\, 10,\, 100$ TeV, respectively.
The effects of one-loop effective Higgs potential are fully included.
The relevant sfermion soft parameters at $M_X$ are assumed to
be given by
$A_{t,b,\tau}/M_{1/2}=-1$,
$m^2_{Q_3,U_3}=(y_t/g_{GUT})^2M_{1/2}^2$,
$m^2_{D_3}=(y_b/g_{GUT})^2M_{1/2}^2$ and
$m^2_{E_3,L_3}=(y_{\tau}/g_{GUT})^2M_{1/2}^2$.
\label{fig:full_rewsb}}
\end{minipage}
\end{center}
\end{figure}
To confirm the above results,
we perform an alternative analysis
including the effects of one-loop effective Higgs
potential \cite{effectivepot,oneloop} as well as the effects
of the $b$-quark and $\tau$-lepton Yukawa couplings.
Here we treat $B\mu$ and $\mu$ at $M_{\rm SB}$ as free parameters
and evaluate their values by minimizing the
one-loop corrected  effective
Higgs potential for fixed values of
$M_Z$, $\tan\beta$ and $|M_{1/2}|$.
We then evolve $B\mu$ and $\mu$ up to $M_X$ and check if the resulting
values  satisfy the relation (\ref{higgs}).
This method works because the RG equations
of other soft SUSY breaking masses
do not depend on $\mu$ or $B\mu$ at one-loop level.
For numerical analysis, we use
$m^{\rm pole}_t=175$ GeV, and
the relevant  sfermion masses and $A$-parameters are assumed to be
$A_{t,b,\tau}/M_{1/2}=-1$,
$m^2_{Q_3,U_3}=(y_t/g_{GUT})^2M_{1/2}^2$,
$m^2_{D_3}=(y_b/g_{GUT})^2M_{1/2}^2$ and
$m^2_{E_3,L_3}=(y_{\tau}/g_{GUT})^2M_{1/2}^2$ at $M_X$,
where $D_3$, $E_3$ and $L_3$ denote the $SU(2)$-singlet
$b$-quark, $SU(2)$-singlet $\tau$-lepton, and
$SU(2)$-doublet $\tau$-lepton superfield, respectively.
For large $\tan\beta$, which is the case that $A_{b,\tau}$
and $m^2_{D_3,E_3,L_3}$ become relevant,
these forms of $A_{b,\tau}$ and $m^2_{D_3,E_3,L_3}$
mimic well  the actual values \cite{future}
for the SUSY breaking by $F^T$ and $F^\Omega$.

The tree-level relations of (\ref{eq:rewsb_relation}) can be rewritten as
\bea
\mu^2 = -\frac{M_Z^2}{2} -\frac{m^2_{H_1}-m^2_{H_2}\tan^2\beta}{1-\tan^2\beta}
&,&~~~|B\mu| = \frac{\tan\beta}{1+\tan^2\beta}(m^2_{H_1}+m^2_{H_2}+2\mu^2).
\eea
Then the effects of one-loop Higgs potential can be included
by replacing $m^2_{H_{1,2}}$
by $m^2_{H_{1,2}} - t_{1,2}/\langle
H^0_{1,2} \rangle$, where $t_{1,2}$ are given by \cite{bbo93}
\bea
t_{1,2} &=& -\frac{1}{32\pi^2}Str\left[\frac{\partial{\cal M}^2}{\partial \langle H^0_{1,2}\rangle}{\cal
 M}^2\left(\ln\frac{{\cal M}^2}{M_{SB}^2}-1\right)\right]\,,
\eea
for ${\cal M}^2$ representing all the mass matrices in the model.
For numerical analysis, we use the form of $t_{1,2}$ given in \cite{pbmz96}.
Because $t_{1,2}$ depend implicitly on $\mu$ through the
masses of neutralino, chargino and squarks, we use numerical
iteration to evaluate $\mu$ starting
from a point around the RG-improved tree-level solution.

In Fig.\ref{fig:full_rewsb}, we show the result of the analysis
computing $m^2_{H_{1,2}}+\mu^2+B\mu$ at $M_X$ as
 a function of $\tan\beta$ for various values of $|M_{1/2}|$.
Because the electroweak symmetry breaking condition
does not depend on the signs of $\mu$ and $B\mu$, we have four
different cases distinguished by the signs of
$\mu$ and $\mu B$.
%For numerical analysis, we use $m^{\rm pole}_t=175$ GeV,
%We
%$A_{t,b,\tau}/M_{1/2}=-1$ and
% $m_{{Q}_3, {U}_3}^2=(y_t/g_{GUT})^2M_{1/2}^2$,
% $m_{{D}_3}^2 =(y_b/g_{GUT})^2M_{1/2}^2$,
%$m_{{E}_3, {L}_3}^2
% =(y_\tau/g_{GUT})^2M_{1/2}^2$  at $M_X$.
%correspond to the
% sign of $\mu$ and two branches in Fig. \ref{fig:tree_rewsb}
%***\footnote{Note that (\ref{prediction1}) breaks the Peccei-Quinn symmetry
% and fixes sign convention of $B\mu$.
%Therefore, in contrast to analyses in the minimal
% supergravity type models,
% sign of $B\mu$ also
% has physical meaning in addition to sign of $B$.}.
Fig.\ref{fig:full_rewsb} shows that
 the condition of gauge-Higgs unification,
i.e. $m^2_{H_{1,2}}+\mu^2+B\mu = 0$ at $M_X$, cannot be satisfied
for reasonable  range of $|M_{1/2}|$.
To satisfy the gauge-Higgs unification condition,
$|M_{1/2}|\gtr 10$ TeV is required
 as we have anticipated from the analysis based on the
RG-improved  tree-level Higgs potential.

\begin{figure}[t]
\begin{center}
\begin{minipage}{12cm}
\centerline{
{\hspace*{-.2cm}\psfig{figure=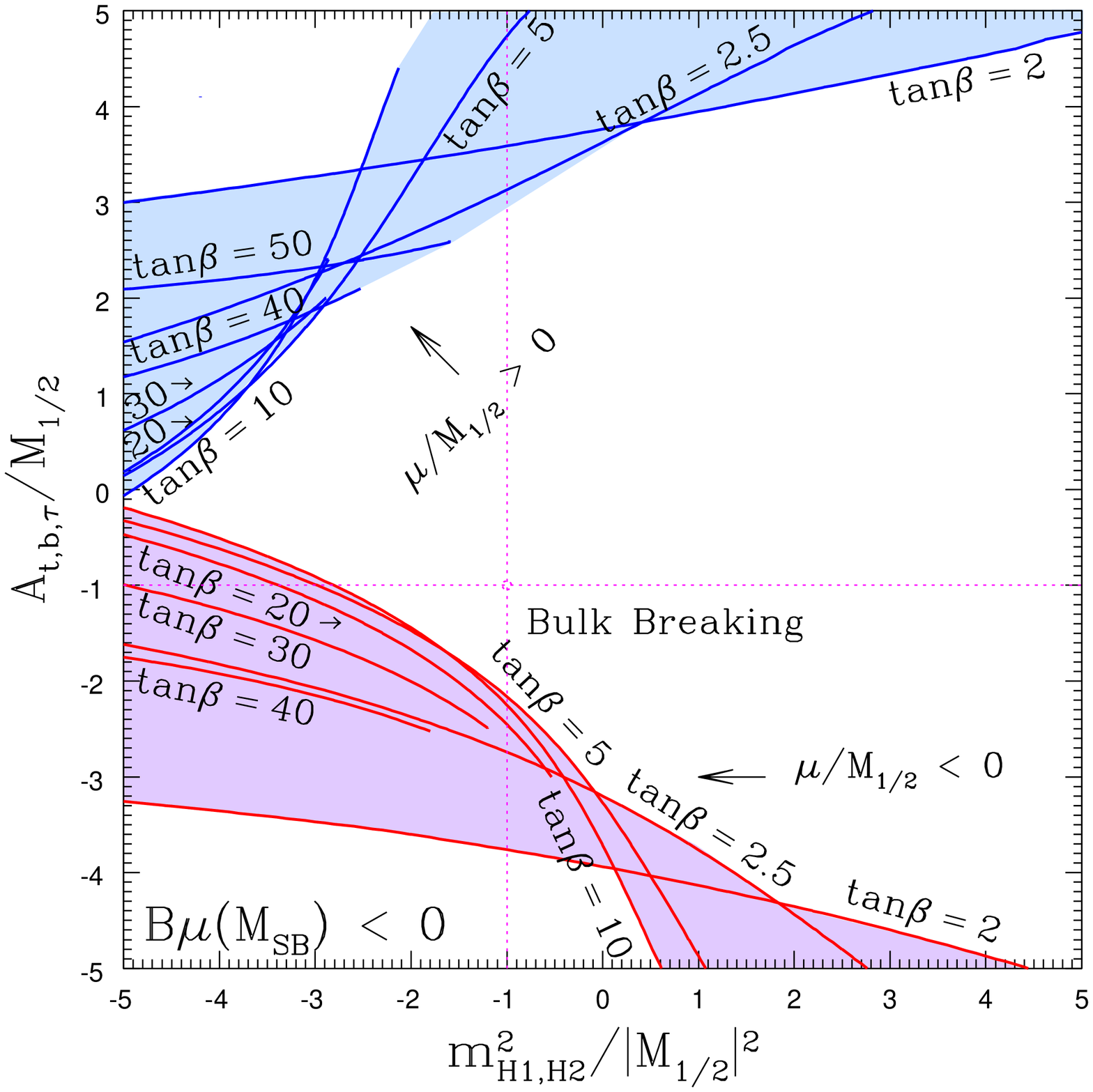,angle=0,width=8.0cm}}
{\hspace*{-.2cm}\psfig{figure=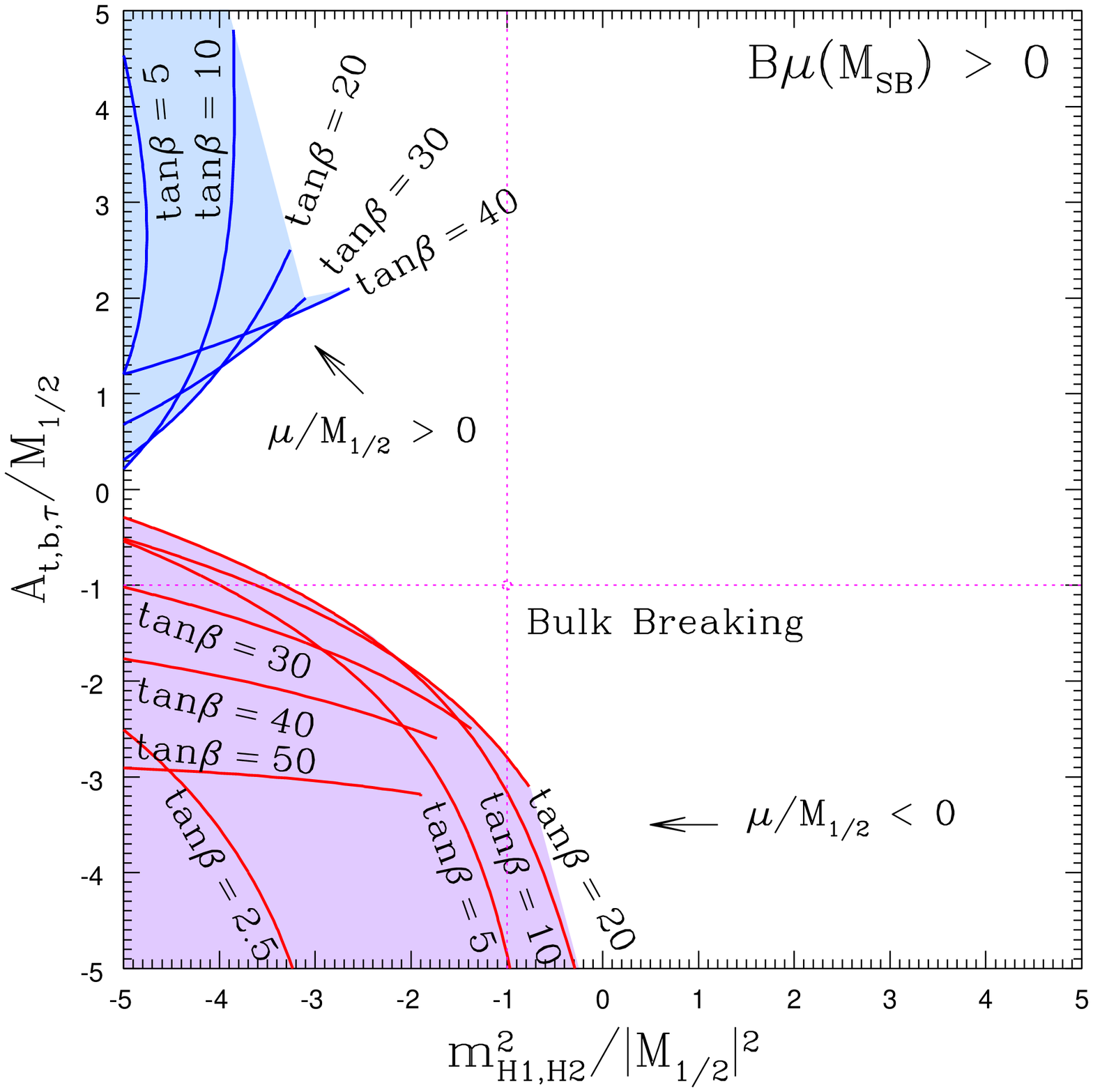,angle=0,width=8.0cm}}
}
\caption{Shaded parameter regions of $A_t/M_{1/2}$ and
$m_{H_1,H_2}^2/M_{1/2}^2$
at $M_X$ give  a correct electroweak symmetry breaking vacuum
when the SUSY breaking effects from the orbifold fixed points
are included for $|M_{1/2}|=500$ GeV.
 \label{fig:general_case}}
\end{minipage}
\end{center}
\end{figure}

\begin{figure}[t]
\begin{center}
\begin{minipage}{12cm}
\centerline{
{\hspace*{-.2cm}\psfig{figure=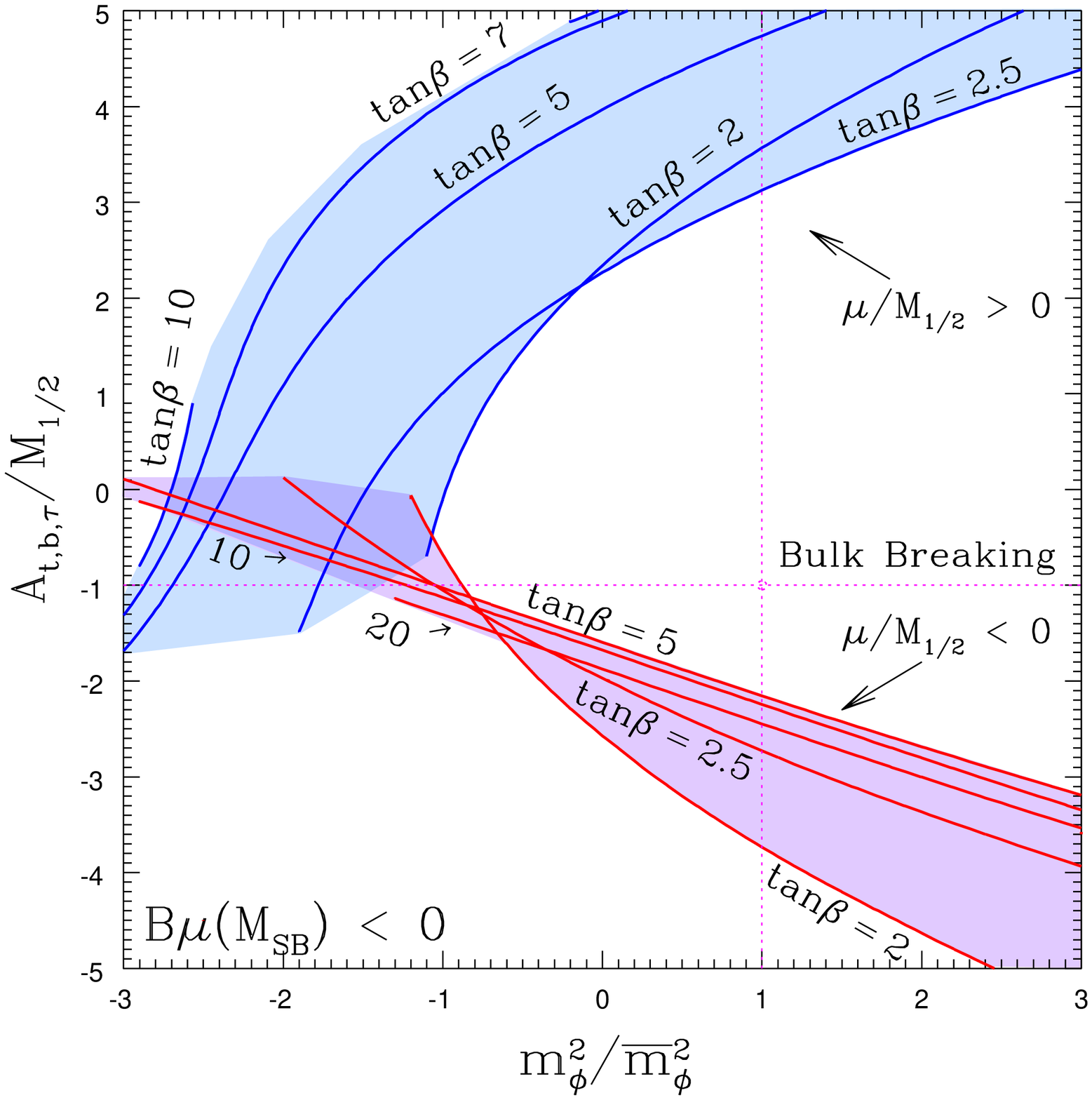,angle=0,width=8.0cm}}
{\hspace*{-.2cm}\psfig{figure=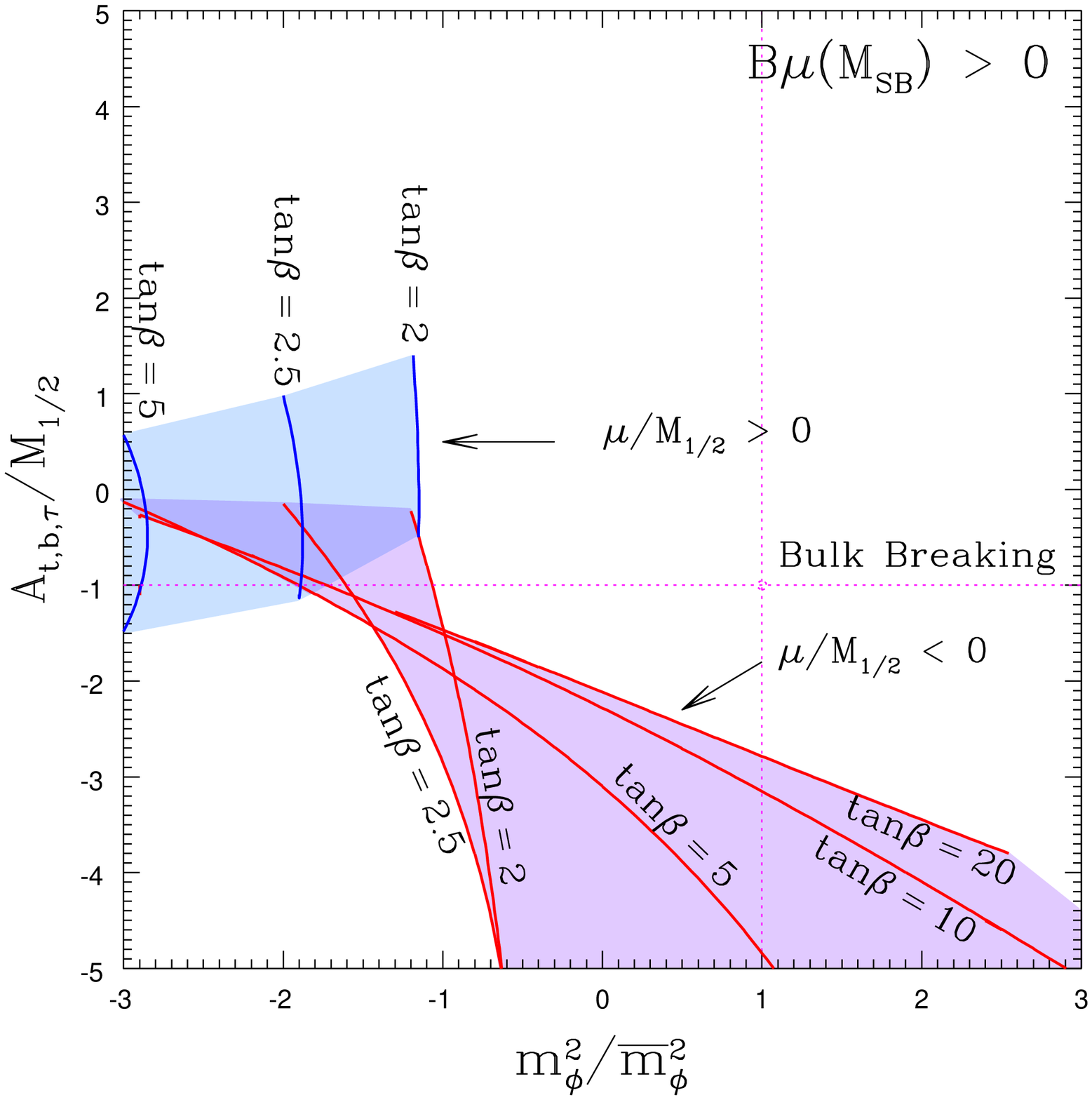,angle=0,width=8.0cm}}
}
\caption{Shaded parameter regions of $A_t/M_{1/2}$ and
$m_\phi^2/\bar{m}^{2}_\phi$ ($\bar{m}^2_\phi=(y_{t,b,\tau}/g_{GUT})^2
M_{1/2}^2$)
at $M_X$ give a  correct electroweak symmetry breaking vacuum
when the SUSY breaking effects from the orbifold fixed points
are included for $|M_{1/2}|=500$ GeV.
 \label{fig:general_case1}}
\end{minipage}
\end{center}
\end{figure}

We saw that a correct electroweak symmetry breaking
is not allowed when SUSY breaking is mediated
dominantly by the bulk $F^T$ and $F^\Omega$
for the reasonable range of $|M_{1/2}|\lesssim 10$ TeV.
This strong constraint is mainly due to the
predictions $m_{H_1,H_2}^2=-M_{1/2}^2$ and
$A_t=-M_{1/2}$ at $M_X$ which are valid for
the SUSY breaking by $F^T$ and $F^\Omega$.
If another source of SUSY breaking is introduced, e.g. the auxiliary
component $F^{\cal Z}$ and/or $F^{{\cal Z}'}$
of ${\cal Z}, {\cal Z}'$ confined at the orbifold fixed point,
these predictions
are not valid anymore, while the prediction (\ref{prediction1})
remains to be valid.
%We examine how much deviation from
%(\ref{higgs}) and (\ref{stop}) is required to
%have a correct electroweak symmetry breaking
%with the relation (\ref{prediction1}).
We examine such general situation
in which  $A_t/M_{1/2}$ and $m^2_{H_1,H_2}/M_{1/2}^2$
at $M_X$ have arbitrary values of order unity, while
the prediction (\ref{prediction1}) is maintained.
For the results
depicted in Fig.\ref{fig:general_case},
we assumed that the
squark and slepton masses at $M_X$ are given by
$m^2_{Q_3,U_3}=(y_t/g_{GUT})^2M_{1/2}^2$,
$m^2_{D_3}=(y_b/g_{GUT})^2M_{1/2}^2$ and
$m^2_{E_3,L_3}=(y_{\tau}/g_{GUT})^2M_{1/2}^2$,
and also $A_t=A_b=A_\tau$.
In fact, the results are not so sensitive
to $m_\phi^2$
($\phi=Q_3,U_3,D_3,L_3,E_3$),
and we obtain
similar results as long as
$m_\phi^2={\cal O}(M_{1/2}^2)$.

The shaded region in Fig.\ref{fig:general_case}
represent the parameter region
yielding a correct electroweak symmetry breaking
for $\tan\beta$ varying from 2 to 60.
Each solid curve indicates a contour of fixed $\tan\beta$.
Fig.\ref{fig:general_case} shows that
if there exists an additional SUSY breaking
 $F^{\cal Z,Z'}={\cal O}(F^T/R)$ or ${\cal O}(F^\Omega)$
from the fixed points,
sizable parameter region of the model can
yield a correct electroweak
symmetry breaking.
In particular, it shows that any value of $\tan\beta>2$
can be realized.
The point $(m^2_H/|M_{1/2}|^2, A_t/M_{1/2})=(-1,-1)$
corresponds to the bulk SUSY breaking by
$F^T$ and $F^\Omega$,
which is obviously outside the shaded regions.
%However, in $B\mu(M_{SB})<0$ case, if we fix $A_0/M_{1/2}=-1$,
% we can achieve correct electroweak symmetry breaking
% with negative extra contribution to $m^2_H$ of
% twice or larger magnitude relative to the Shark-Shwarz breaking.
%Alternatively, if we fix $m^2_H/|M_{1/2}|^2=-1$, additional contribution
%to the A-parameter of $A_0/M_{1/2}\gtr 3$ for $\mu/M_{1/2}>0$
% or $A_0/M_{1/2}\ler -2$ for $\mu/M_{1/2}<0$ can satisfy the condition.
%In $B\mu(M_{SB})>0$ case, we have allowed region with tighter bounds
% as $m^2_H/|M_{1/2}|^2\ler-3$ or
% $A_0/M_{1/2}\ler-3$ only for $\mu/M_{1/2}<0 $ respectively.

We also performed the analysis for
the case that $A_t/M_{1/2}$ and $m_\phi^2/M_{1/2}^2$
at $M_X$
take arbitrary values of order unity, while
$m^2_{H_1,H_2}/M_{1/2}^2=-1$.
The results are depicted in
  Fig.\ref{fig:general_case1}.
Here $\bar{m}^{2}_{\phi}$ is defined
as $(y_{t,b,\tau}/g_{GUT})^2\,M^2_{1/2}$, and other parameters are same as
in Fig.3\,, so
the point $(m^2_{\phi}/\bar{m}^{2}_{\phi}, A_t/M_{1/2})=(1,-1)$
corresponds to the SUSY breaking by
the bulk $F^T$ and $F^\Omega$.
Fig.\ref{fig:general_case1}
 shows that the results
 are not so sensitive to $m^2_{\phi}$
as long as $m_\phi^2$ at $M_X$ is positive.
It shows also that to achieve a correct electroweak
 symmetry breaking with $A_t\,\approx\,-M_{1/2}$,
a negative stop mass
 of ${\cal O}(M^2_{1/2})$ is required at $M_X$,
which becomes positive at $M_{SB}$ due to the RG evolution.

\section{conclusion}

In this paper, we have examined the Higgs mass parameters
and electroweak symmetry breaking in supersymmetric orbifold
field theories in which the 4D Higgs fields originate
from higher-dimensional gauge supermultiplets.
To be specific, we focused on 5D models, however some of
our results are more generic.
It is noted that the gauge-Higgs unification
within orbifold field theory leads to a specific boundary
condition on the Higgs mass parameters
at the compactification scale $M_X$,
which has been obtained also in supersymmetric pseudo-Goldstone
Higgs models.
More restrictive boundary conditions on the
Higgs mass parameters could be obtained
when SUSY breaking is mediated dominantly by
the auxiliary components of the radion superfield $T$ and
the 4D SUGRA multiplet, which corresponds to the Sherk-Schwarz
SUSY breaking with additional SUSY breaking from
the conformal compensator superfield $\Omega$.
If the 4D effective theory at scales below $M_X$ is the MSSM,
the SUSY breaking by $F^T$ and $F^\Omega$ alone can not
give a correct electroweak symmetry breaking vacuum
for reasonable range of parameters.
So we need additional
SUSY breaking mediated for instance by a brane superfield
${\cal Z}$ confined at the orbifold fixed points.
If $F^{\cal Z}$ is included, there exist a sizable
portion of parameter space which can give correct electroweak
symmetry breaking with wide range of $\tan\beta$.
The results of numerical analysis are summarized
in Fig.1-Fig.4.

\bigskip
{\bf Acknowledgements}

\medskip

This work is supported by KRF PBRG 2002-070-C00022, the Grants-in-aid
from the Ministry of Education, Culture, Sports, Science and Technology, Japan,
No.12046201, 14039207, 14046208, 14740164
and the 21st century COE program, "Exploring New Science by 
Bridging Particle-Matter Hierarchy".

\end{document}